\documentclass{CS-ieee-sssm}

\newcommand{\SetR}{{\rm I\!R }}

\title{Portfolio Selection Under Buy-In Threshold Constraints Using DC Programming and DCA\thanks{1-4244-0451-7/06/\$20.00 \copyright2006 IEEE}}

\author{Hoai An LE THI\inst{1}, Mahdi MOEINI\inst{2}}

\affiliation{\inst{1}
Universit\'e Paul Verlaine - Metz, France (lethi@univ-metz.fr)  \\
\inst{2}Universit\'e Paul Verlaine - Metz, France
(moeini@univ-metz.fr)}

\abstract{In matter of Portfolio selection,  we consider a
generalization of the Markowitz Mean-Variance model which includes
\emph{buy-in threshold constraints}. These constraints limit the
amount of capital to be invested in each asset and prevent very
small investments in any asset. The new model can be converted
into a NP-hard mixed integer quadratic programming problem. The
purpose of this paper is to investigate a continuous approach
based on DC programming and DCA (DC Algorithms) for solving this
new model. DCA  is a local continuous approach to solve a wide
variety of nonconvex programs for which it provided quite often a
global solution and proved to be more robust and efficient than
standard methods. Preliminary comparative results of DCA and a
classical Branch-and-Bound algorithm will be presented. These
results show that DCA is an efficient and promising approach for
the considered portfolio selection problem.}


\begin{document}
\maketitle
$$
$$
$$
$$
$$
$$

\section{Introduction}
 In the portfolio selection problem, given a set of
available securities or assets, we want to find out the optimum
way of investing a particular amount of money in these assets.
Each way of the different ways to diversify this money between the
several assets is called a portfolio \cite{neural2006}. For
solving this portfolio problem, Markowitz \cite{markowitz52,
markowitz59} has set up a quantitative frame-work. Markowitz's
model which is called Mean-Variance model assumes that the return
on a portfolio of assets can be completely described by the
expected return and the variance of returns (risk) between these
assets. For a particular universe of assets, the set of portfolios
of assets that offer the minimum risk for a given level of return
is the set of efficient portfolios. These portfolios can be found
by convex quadratic programs (QP). But the Markowitz's standard
model, does not contain some practical constraints. For example,
the standard Mean-Variance model has not got any bounding
constraints limiting the amount of money to be invested in each
asset neither prevents very small amounts of investments in each
asset. This kind of constraints is very useful in practice and is
called {\it buy-in threshold constraints} \cite{bartholomew2005}.
In order to overcome these inconveniences, the standard model can
be generalized to include these constraints.

In this paper we focus on solving the problem of portfolio
selection under buy-in threshold constraints. We investigate a
local deterministic approach based on DC (Difference of Convex
functions) programming and DCA (DC Algorithms) that were
introduced by Pham Dinh Tao in their preliminary form in 1985.
They have been extensively developed since 1994 by Le Thi Hoai An
and Pham Dinh Tao and become now classic and more and more popular
(see e.g. \cite{Harring05}, \cite{lethithesis} - \cite{lethi2005},
\cite {PLT97, PLT98}, \cite{WSSH04} and references therein). DCA
has been successfully applied to many large-scale (smooth or
nonsmooth) nonconvex programs in various domains of applied
sciences,  for which it provided quite often a global solution and
proved to be more robust and efficient than standard methods (see
e.g. \cite{Harring05}, \cite{lethithesis} - \cite{lethi2005},
\cite {PLT97,PLT98},    \cite{WSSH04} and reference therein).

$$
$$
$$
$$
$$
$$
$$
$$

The existence of buy-in threshold constraints makes the
corresponding portfolio selection problem nonconvex and so very
difficult to solve by existing algorithms. By introducing the
binary variables, we first express the buy-in threshold
constraints as mixed zero-one linear constraints; then, using an
exact penalty result, we reformulate the last problem in terms of
a DC program. A so-called DC  program is that of minimizing a DC
function over a convex set. We then suggested using DC programming
approach and DCA to solve this portfolio selection problem. For
testing the efficiency of DCA we compare it with a
Branch-and-Bound algorithm.

The paper is organized as follows. After the introduction, we present  in section
2  the model of the portfolio selection
problem under buy-in threshold constraints, and the reformulation
in term of a DC program. Section 3 deals with DC programming and
a special realization of DCA to the underlying portfolio problem.
Section 4 is
devoted to preliminary experimental results and some conclusions are
reported in section 5.


\section{Portfolio selection  problem under buy-in threshold constraints}
\subsection{Problem formulation}

First of all, as we introduce the notations that we are going to
use in this paper, let us remind the well known Markowitz's
\cite{markowitz52, markowitz59} Mean-Variance model for the
portfolio selection problem. Let $n$ be the number of available
assets, $r_{i}$ be the mean return of asset $i$, $Q$ be an
$n\times n$ Variance-Covariance (positive semidefinite) matrix
such that its $(i,j)$-th element, that is $\sigma_{i,j}$ is the
covariance between returns of assets $i$ and $j$ and its value is
calculated by using the following formula:
\begin{equation}
\sigma_{ij}=(1/m)\sum\limits_{k=1}^{m}((r_{ik}-r_{i})(r_{jk}-r_{j})).
\end{equation}
Here $r_{ik}$ is the $(i,k)$-th historical data and $m$ is the
number of periods that we have considered. Let $R$ be the desired
expected return and the decision variables $y_{i}$ represent the
proportion $(0\leq y_{i} \leq 1)$ of capital to be invested in
asset $i$ and $y^{T}=(y_{1}, ... , y_{n})$. Using this notation,
the standard Markowitz's Mean-Variance model
is (\cite{bartholomew2005})

\begin{equation}
\min {V(y)}:=y^{T}Qy
\end{equation}
\begin{center}
$ s.t:$ $
\left\{\sum\limits_{i=1}^{n}r_{i}y_{i}=R, \sum\limits_{i=1}^{n}y_{i}=1,
y_{i}\geq 0, \:  i = 1, \ldots, n\right\}.
$
\end{center}
By solving this problem, one minimizes the total variance (risk)
associated with the portfolio by ensuring that the portfolio has
an expected return $R$. In this paper no short-sale is allowed.\\
This formulation is a simple   convex quadratic
program for which efficient algorithms
are available. By resolving the above QP for varying values of $R$, we can
trace out the efficient frontier, a smooth non-decreasing curve
that gives the best possible tradeoff  of risk against return.

For generalizing the standard Markowitz model with the inclusion of
  \emph{buy-in threshold} constraints, we will use some
additional notations. Let $a_{i}$ and $b_{i}$ be, respectively,
the lower and upper bounds for the proportion of capital to be
invested in asset $i$, with $0< a_{i} \leq b_{i} \leq 1$. The
generalized Mean-Variance model for the portfolio selection
problem under buy-in threshold constraints can be written as
\begin{equation}
\min {V(y)}:=y^{T}Qy
\end{equation}
 s.t:
$$\left\{\sum\limits_{i=1}^{n}r_{i}y_{i}=R, \sum\limits_{i=1}^{n}y_{i}=1,
y_{i}\in \{0\} \cup [a_{i}, b_{i}], \:  i = 1, \ldots, n\right\}.
$$
Due to the last constraints $y_{i}\in \{0\} \cup [a_{i}, b_{i}]$,
this is a hard problem
for which efficient algorithms are
not available.

\subsection{Reformulation}
The later problem can be reformulated as a mixed integer
quadratic problem by introducing the additional variables $z_{i}$
such that
$$z_{i}=1 \mbox{ iff $y_i \in [a_{i}, b_{i}]$, $0$ otherwise.}$$
 The new
mixed integer   quadratic programming formulation of the problem
is
\begin{equation}
\label{eq4}
\min {V(y)}:=y^{T}Qy
\end{equation}
$$s.t: \quad
\sum\limits_{i=1}^{n}r_{i}y_{i}=R, \sum\limits_{i=1}^{n}y_{i}=1,$$
$$a_{i}z_{i} \leq y_{i} \leq b_{i}z_{i}, z_{i}\in
\{0,1\}, \quad i = 1, \ldots, n .
$$

Using the exact penalty result presented in \cite{lethi2005b}, we will
formulate (\ref{eq4}) in the form of a convex-concave minimization problem with
linear constraints which is consequently  a DC program. Let
$\begin{array}{cc}
 {A} := &  \{(y,z)\in \SetR^{n} \times
[0,1]^{n}:\sum\limits_{i=1}^{n}r_{i}y_{i}=R,\\
 & \sum\limits_{i=1}^{n}y_{i}=1, a_{i}z_{i} \leq y_{i} \leq
b_{i}z_{i},  \quad i = 1, \ldots, n\}
\end{array}
$.

Define the function
$$ p(y,z):= \sum\limits_{i=1}^{n}z_{i}(1-z_{i}). $$
Clearly, $p$ is a concave function with nonnegative values on $A$ and
the feasible region of (\ref{eq4}) can be written as
$$\{(y,z)\in A: z_{i} \in \{0,1\} \}$$
$$= \{(y,z)\in A: p(y,z)=0 \} =
\{(y,z)\in A: p(y,z)\leq 0 \}. $$
So, (\ref{eq4}) can be expressed as
\begin{equation}
\label{eq5}
\min \{ V(y):=y^{T}Qy :
(y,z)\in A, \quad  p(y,z)\leq 0\}.
\end{equation}
Since the objective function $V$ is convex and $A$ is a bounded polyhedral convex set,
according to
\cite{lethi2005b}, there is $t_{0}\geq 0$ such that for any $t>
t_{0}$, the program (\ref{eq5}) is equivalent to
\begin{equation}
\label{eq6}
\min \{F(y,z) :=y^{T}Qy+tp(y,z) : (y,z) \in A \}.
\end{equation}

The function $F$ is convex in variable $y$ and concave in variable $z$.
Consequently it is  a DC function.
A natural DC formulation of the problem (\ref{eq6}) is
 $$ \min \{g(y,z)-h(y,z):(y,z)\in \SetR^{n}\times \SetR^{n} \},$$
where
$$ g(y,z):=y^{T}Qy+\chi _{A}(y,z), $$
 and
$$ h(y,z):=t\sum\limits_{i=1}^{n}z_{i}(z_{i}-1). $$
Here $\chi _{A}$ is the indicator function on $A$, i.e. $\chi
_{A}(y,z)=0$ if $(y,z)\in A$ and $+\infty$ otherwise.


\section{Solution method via DC programming and DCA}

\subsection{DCA for general DC programs}
Let $\Gamma _{0}(%
\mathrm{I\!R}^{n})$ denote the convex cone of all lower semicontinuous
proper convex functions  on $\SetR^{n}$, and consider the
general DC program
\begin{equation}
(P_{dc})\quad \alpha = \inf\{f(x):=g(x)-h(x):x\in \SetR^{n}\}
\end{equation}
where $g,h\in \Gamma _{0}(\SetR^{n})$. Such a function $f$
is called DC function, and $g-h$, DC~decomposition of $f$ while the convex
functions $g$ and $h$ are DC components of $f.$ \newline
 Let $C$ be a convex set. The problem
\begin{equation}
\inf\{f(x):=k(x)-h(x):x\in C\}
\end{equation}
can be transformed to an unconstrained DC program by using the
indicator function on $C$, i.e.,
\begin{equation}
\inf\{f(x):=g(x)-h(x):x\in \SetR^{n}\}
\end{equation}
where $g:=k+\chi _{C}$.\\
Let $g^{\ast}(y):= \sup\{\langle x,y \rangle-g(x) : x\in \SetR^{n}\}$ be the
conjugate function of $g$. Then, the following program is called
the dual program of (P$_{dc}$):
\begin{equation}
(D_{dc}) \quad \alpha_D = \inf\{h^{\ast} (y)-g^{\ast} (y):y\in \SetR^{n}\}.
\end{equation}
Under the natural convention
in DC~programming that is $+ \infty -(+\infty)= + \infty $, and by using the fact that every function $%
h\in $ $\Gamma _{0}(\mathrm{I\!R}^{n})$ is characterized as a pointwise
supremum of a collection of affine functions, say
\[
h(x):=\sup \{\langle x,y\rangle -h^{\ast }(y):y\in \mathrm{I\!R}^{n}\},
\]
one can prove that $\alpha = \alpha_D$.
We observe the perfect symmetry between primal and dual DC~programs: the
dual to $(D_{dc})$ is exactly $(P_{dc}).$\\
Recall that, for $\theta \in $ $\Gamma _{0}(\mathrm{I\!R}^{n})$ and $%
x_{0}\in dom$ $\theta :=\{x\in \mathrm{I\!R}^{n}: \theta (x_{0})<+\infty \},$
$\partial \theta (x_{0})$ denotes the subdifferential of $\theta $ at $%
x_{0},$ i.e., (\cite{Rockafellar})
\begin{equation}
\partial \theta (x_{0}):=\{y\in \mathrm{I\!R}^{n}: \theta (x)\geq \theta
(x_{0})+\langle x-x_{0},y\rangle ,\forall x\in \mathrm{I\!R}^{n}\}.
\label{subdif}
\end{equation}
The subdifferential $\partial \theta (x_{0})$ is a closed convex set in $%
\mathrm{I\!R}^{n}.$ It generalizes the derivative in the sense that $\theta $
is differentiable at $x_{0}$ if and only if $\partial \theta (x_{0})$ is
reduced to a singleton which is exactly$\{\theta ^{\prime }(x_{0})\}.$
The necessary local optimality condition for the primal DC program
(P$_{dc}$) is:
\begin{equation}
\label{local}
\partial g(x^\ast) \supset \partial h(x^\ast).
\end{equation}
A point $x^*$ verifies the  condition $\partial h(x^\ast) \cap \partial g(x^\ast)\neq
\emptyset$ is called a critical point of $g-h$.
The condition (\ref{local}) is also sufficient for many important classes of
DC programs, for example, in case of the function $f$ \ is locally convex at $x^{\ast }$ (\cite{lethi2001,lethi2005,PLT97}).

The transportation of global solutions between $(P_{dc})$ and $(D_{dc})$
is expressed by:
\begin{equation}
 \label{eqiii}
\lbrack \cup _{y^{\ast }\in \mathcal{D}}\,\partial g^{\ast }(y^{\ast
})]\subset \mathcal{P}, \quad [\cup _{x^{\ast }\in \mathcal{P}}\,\partial
h(x^{\ast })]\subset \mathcal{D}
\end{equation}
where\ $\mathcal{P}$ and $\mathcal{D}$ denote the solution sets of $(P_{dc})$
and $(D_{dc})$ respectively. Under technical conditions, this transportation
holds also for local solutions of $(P_{dc})$ and $(D_{dc})$ (\cite{lethithesis, lethi2005,PLT97,PLT98}).

Based on local optimality conditions and duality in DC programming, the DCA
consists in the construction of two sequences $\{x^{k}\}$ and $\{y^{k}\}$,
candidates to be optimal solutions of primal and dual programs respectively,
such that the sequences $\{g(x^{k})-h(x^{k})\}$ and $\{h^{\ast
}(y^{k})-g^{\ast }(y^{k})\}\ $ are decreasing, and $\{x^{k}\}$ (resp. $%
\{y^{k}\}$) converges to a primal feasible solution $\widetilde{x}$ (resp. a
dual feasible solution $\widetilde{y}$) verifying local optimality
conditions and
\begin{equation}
\widetilde{x}\in \partial g^{\ast }(\widetilde{y}),\quad \widetilde{y}\in
\partial h(\widetilde{x}).  \label{intersubdif}
\end{equation}
The DCA then yields the next scheme:
\begin{equation}
y^{k}\in \partial h(x^{k});\quad x^{k+1}\in \partial g^{\ast }(y^{k}).
\label{DCAscheme}
\end{equation}
In other words, these two sequences $\{x^{k}\}$\ and $\{y^{k}\}$\ are determined in the way
that $x^{k+1}$\ (resp. $y^{k}$) is a solution to the convex program $(P_{k})$
(resp. $(D_{k}))$ defined by
\[
\inf \{g(x)-h(x^{k})-\langle x-x^{k},y^{k}\rangle :x\in \mathrm{I\!R}^{n}{\},%
}\quad (P_{k})
\]
\[
\inf \{h^{\ast }(y)-g^{\ast }(y^{k-1})-\langle y-y^{k-1},x^{k}\rangle :y\in
\mathrm{I\!R}^{n}{\}}\quad (D_{k}).
\]
In fact, at each iteration one
replaces in the primal DC program $(P_{dc})$ the second component $h$ by its
affine minorization $h_{k}(x):=h(x^{k})+\langle x-x^{k},y^{k}\rangle $ at a
neighbourhood of $x^{k}$ to give birth to the convex program $(P_{k})$ whose
the solution set is nothing but $\partial g^{\ast }(y^{k}).$ Likewise, the
second DC~component $g^{\ast }$ of the dual DC program $(D_{dc})$ is
replaced by its affine minorization $(g^{\ast })_{k}(y):=g^{\ast
}(y^{k})+\langle y-y^{k},x^{k+1}\rangle $ at a neighbourhood of $y^{k}$ to
obtain the convex program $(D_{k})$ whose $\partial h(x^{k+1})$ is the
solution set. DCA performs so a double linearization with the help of the
subgradients of $h$ and $g^{\ast }$.

It is worth noting that (\cite{lethithesis,lethi2005,PLT97,PLT98}
 DCA works with the convex DC
components $g$ and $h$ but not the DC function $f$ \ itself. Moreover, a DC
function $f$ \textit{has infinitely many DC decompositions which have
crucial impacts on the qualities} (speed of convergence, robustness,
efficiency, globality of computed solutions,...) of DCA.

Convergence properties of DCA and its theoretical basis can be found in \cite{lethithesis,lethi2005,
PLT97}, for instant it is important to mention that
\begin{itemize}
\item DCA is a descent method (the sequences\ $\{g(x^{k})-h(x^{k})\}$\ and\ $\{h^{\ast
}(y^{k})-g^{\ast }(y^{k})\}$\ are\ decreasing) without linesearch;
\item If the optimal value $\alpha \ $of problem $(P_{dc})$ is
finite and the infinite sequences $\{x^{k}\}\ $and $\{y^{k}\}$\ are bounded
then every limit point $\widetilde{x}\ $(resp. $\widetilde{y}$) of the
sequence $\{x^{k}\}$\ (resp. $\{y^{k}\})$\ is a critical point of $g$ $-$ $h$%
\ (resp. $h^{\ast }-g^{\ast }$).
\item DCA has a linear convergence for \ general DC programs.
\end{itemize}
\subsection{DCA for solving (\ref{eq6})}
According to the general framework of DCA, we first need computing
a sub-gradient of the function $h$ defined by
$h(y,z):=t\sum\limits_{i=1}^{n}z_{i}(1-z_{i})$.
 From the definition of $h$ we have
\begin{equation}
(u^{k},v^{k})\in \partial h(y^{k},z^{k})\Leftrightarrow
u^{k}_{i}=0  , v^{k}_{j}=t(2z^{k}_{j}-1),
\end{equation}
$$i,j=1,\ldots,n.$$
Secondly, we have to compute an optimal solution of the following
convex quadratic program
\begin{equation}
\min \{y^{T}Qy- \langle(y,z),(u^{k},v^{k})\rangle :(y,z)\in A\}
\end{equation}
that will be $(y^{k+1},z^{k+1})$.
To sum up, the DCA applied to (\ref{eq6}) can be described as follows.

\noindent \textbf{Algorithm DCA}
\begin{enumerate}
    \item \textbf{Initialization}: Let $\varepsilon$
    be a sufficiently small positive number, let
    $(y^{0},z^{0})\in R^{n}\times [0,1]^{n}$, and set $k=0$;
    \item \textbf{Iteration}: $k=0,1,2,...$\\
    set $u^{k}_{i}=0$ and $v^{k}_{i}=t(2z^{k}_{i}-1)$ for $i=1,...,n$.\\
    Solve the following quadratic program
    \begin{equation}
    \min \{y^{T}Qy-\langle(y,z),(u^{k},v^{k})\rangle :(y,z)\in A\}
    \end{equation}
    to obtain $(y^{k+1},z^{k+1})$.\\
    \item If $\parallel y^{k+1}-y^{k}\parallel + \parallel z^{k+1}-z^{k}\parallel \leq \varepsilon
    $, then stop, $(y^{k+1},z^{k+1})$ is a solution, otherwise set
    $k=k+1$ and go to step 2.
\end{enumerate}
For evaluating the quality of solutions computed by DCA and by the way their globality,  we solve the problem by a classical Branch-and-Bound
algorithm for mixed zero-one programming (\ref{eq4}). More precisely,
the lower bound is computed by solving the classical relaxed problem of
(\ref{eq4})
(the binary constraints $z_{i} \in \{0,1\}$ are replaced by $0\leq z_{i} \leq 1$) which is
a convex quadratic program, and the upper bound is updated when
a better feasible solution to (\ref{eq4}) is discovered.
The subdivision is performed in the way that $z_{i} = 0$ or
 $z_{i} = 1$.


\section{Computational experiments}
 We have tested the algorithms on two sets
of data that have been already used in \cite{beasley2000,
neural2006, mitra2001}. These data correspond to weekly prices
from March 1992 to September 1997 and they come from the indices :
Dax 100 in Germany and Nikkei 225 in Japan. The number $n$ of
different assets considered for each one of the test problems is
$85$ and $225$, respectively. The mean returns and covariances
between these returns have been calculated for the data. All the
results presented here have been computed using the values
$a_{i}=0.05$ and $b_{i}=1.0$ in (\ref{eq4}). We have tested  DCA
and the classical Branch-and-Bound algorithm  for different values
of desired expected return $R$.  The parameter $t$ is taken the
value $0.01$ for the first set of data and $0.02$ for the second
one. The tolerance $\varepsilon$ is equal to $10^{-7}$.
 \\
The algorithms are coded in C++ and run on
 a Pentium $1.600GHz$ of $512$ DDRAM.

\noindent{\bf Finding a good initial point for DCA.}

In fact, one of the key
questions in DCA is    how to find a good
initial solution for it. The question is still open.
In this work, in order to find a
good initial solution we first solve the relaxed problem of  (\ref{eq4}). In general the obtained solution
is not necessarily integer and thus we have to modify it to get
a feasible solution  to (\ref{eq4}). This new solution is taken as the initial point
for DCA.
 The procedure can be
summarized   as follows:
\begin{enumerate}
    \item \textbf{Solution of the relaxed problem}\\ Solve the
    relaxed problem of (\ref{eq4}) to obtain the optimal solution
    $(\widetilde{y},\widetilde{z})$.
    \item \textbf{Finding an integer solution}\\ obtain an integer
    solution $\widehat{z}$ by rounding each nonzero value
    $\widetilde{z}_i$ to one.
\end{enumerate}
The new solution $(\widetilde{y},\widehat{z})$ may not be   feasible to
(\ref{eq6}).
We need just one iteration of DCA to obtain a feasible solution
 of   (\ref{eq6}), and all the other iterations of
DCA will improve the solution.

We have tested DCA from different initial points:
\begin{itemize}
\item  The point obtained by the above procedure;
\item The optimal solution of the  relaxed problem of (\ref{eq4});
\item The optimal solution of the next problem
$$\min \left\{p(y,z):=  \sum\limits_{i=1}^{n}z_{i}(1-z_{i}) : (y,z) \in A\right\}.$$
\end{itemize}
In our experiments the initial point of DCA given by the first procedure is
the best.

In Tables 1, 2, 3, and 4, we give the results for two considered
data sets. In these tables, the number of iterations (iter),
the computer time in seconds (CPU),
and the solutions obtained by each of the
algorithms are shown. \\
\begin{table}[!ht]
\caption{Numerical results of Branch-and-Bound algorithm for the first
set of data}
\begin{center}
\begin{tabular}{|c|c|c|c|c|}\hline
               R        & Optimal value &  iter & CPU \\
               \hline \hline
               0.00001  & 0.000305      & 12  & 10.953  \\ \hline
               0.00002  & 0.000305      & 13  & 10.656  \\ \hline
               0.00003  & 0.000305      & 12  & 10.516  \\ \hline
               0.00004  & 0.000305      & 12  & 11.703  \\ \hline
               0.00005  & 0.000305      & 13  & 11.610  \\ \hline
               0.00006  & 0.000305      & 13  & 11.547  \\ \hline
               0.00007  & 0.000305      & 13  & 11.672  \\ \hline
               0.00008  & 0.000305      & 13  & 11.813  \\ \hline
               0.00009  & 0.000305      & 13  & 11.813  \\ \hline
               0.0001   & 0.000305      & 13  & 11.890  \\ \hline
               0.0002   & 0.000305      & 14  & 13.110  \\ \hline
               0.0003   & 0.000306      & 14  & 13.110  \\ \hline
               0.0004   & 0.000308      & 16  & 15.407  \\ \hline
               0.0005   & 0.000310      & 24  & 22.844   \\ \hline
               0.0006   & 0.000312      & 15  & 14.250  \\ \hline
               0.0007   & 0.000315      & 15  & 14.328  \\ \hline
               0.0008   & 0.000319      & 32  & 30.563  \\ \hline
               0.0009   & 0.000322      & 32  & 30.563  \\ \hline
               0.001    & 0.000326      & 30  & 29.265  \\ \hline
               0.002    & 0.000390      & 12  & 12.140  \\ \hline
               0.003    & 0.000517      & 11  & 11.657  \\ \hline
\end{tabular}
\end{center}
\label{table1}
\end{table}
\begin{table}[!ht]
\caption{Numerical results of DCA  for the first set of data}
\begin{center}
\begin{tabular}{|c|c|c|c|c|}\hline
               R        & Optimal value & iter& CPU \\
               \hline \hline
               0.00001  & 0.000306      & 2  & 1.594  \\ \hline
               0.00002  & 0.000306      & 2  & 1.609  \\ \hline
               0.00003  & 0.000306      & 2  & 1.594  \\ \hline
               0.00004  & 0.000306      & 2  & 1.625  \\ \hline
               0.00005  & 0.000306      & 2  & 1.609  \\ \hline
               0.00006  & 0.000306      & 2  & 1.578  \\ \hline
               0.00007  & 0.000306      & 2  & 1.610  \\ \hline
               0.00008  & 0.000306      & 2  & 1.594  \\ \hline
               0.00009  & 0.000306      & 2  & 1.609  \\ \hline
               0.0001   & 0.000306      & 2  & 1.703  \\ \hline
               0.0002   & 0.000305      & 2  & 1.750  \\ \hline
               0.0003   & 0.000307      & 2  & 1.719  \\ \hline
               0.0004   & 0.000310      & 2  & 1.781  \\ \hline
               0.0005   & 0.000311      & 2  & 1.735  \\ \hline
               0.0006   & 0.000314      & 2  & 1.719  \\ \hline
               0.0007   & 0.000316      & 2  & 1.719  \\ \hline
               0.0008   & 0.000322      & 2  & 1.781  \\ \hline
               0.0009   & 0.000324      & 2  & 1.687  \\ \hline
               0.001    & 0.000328      & 2  & 1.781  \\ \hline
               0.002    & 0.000391      & 2  & 1.828  \\ \hline
               0.003    & 0.000519      & 2  & 1.953  \\ \hline
\end{tabular}
\end{center}
\label{table2}
\end{table}

\begin{table}[!ht]
\caption{Numerical results of Branch-and-Bound algorithm  for the
second set of data}
\begin{center}
\begin{tabular}{|c|c|c|c|c|}\hline
               R        & Optimal value & iter & CPU \\
               \hline \hline
               0.0001   & 0.000174      & 1348  & 147.953  \\ \hline
               0.0002   & 0.000170      & 718   & 77.343   \\ \hline
               0.0003   & 0.000167      & 491   & 53.328   \\ \hline
               0.0004   & 0.000164      & 549   & 59.313   \\ \hline
               0.0005   & 0.000162      & 671   & 72.625   \\ \hline
               0.0006   & 0.000159      & 788   & 86.500   \\ \hline
               0.0007   & 0.000158      & 1475  & 158.547  \\ \hline
               0.0008   & 0.000156      & 1648  & 175.828  \\ \hline
               0.0009   & 0.000154      & 1838  & 194.860  \\ \hline
               0.001    & 0.000153      & 1980  & 209.610  \\ \hline
               0.002    & 0.000141      & 204   & 22.062   \\ \hline
               0.003    & 0.000147      & 140   & 15.875   \\ \hline
               0.004    & 0.000170      & 129   & 14.406  \\ \hline
\end{tabular}
\end{center}
\label{table1}
\end{table}

\begin{table}[!ht]
\caption{Numerical results of DCA  for the second set of
data}
\begin{center}
\begin{tabular}{|c|c|c|c|c|}\hline
               R        & Optimal value & iter & CPU \\
               \hline \hline
               0.0001   & 0.000186      & 2   & 0.235   \\ \hline
               0.0002   & 0.000189      & 2   & 0.234  \\ \hline
               0.0003   & 0.000193      & 2   & 0.218   \\ \hline
               0.0004   & 0.000182      & 3   & 0.266  \\ \hline
               0.0005   & 0.000174      & 3   & 0.266   \\ \hline
               0.0006   & 0.000173      & 4   & 0.312   \\ \hline
               0.0007   & 0.000170      & 4   & 0.313   \\ \hline
               0.0008   & 0.000167      & 3   & 0.266  \\ \hline
               0.0009   & 0.000167      & 4   & 0.313   \\ \hline
               0.001    & 0.000167      & 4   & 0.312   \\ \hline
               0.002    & 0.000156      & 2   & 0.219  \\ \hline
               0.003    & 0.000159      & 2   & 0.234  \\ \hline
               0.004    & 0.000207      & 2   & 0.203 \\ \hline
\end{tabular}
\end{center}
\label{table1}
\end{table}

The computational results   show that DCA gives a good
approximation of the optimal solution within a very short time.
The running time is less than 2 seconds and the number of iterations is
at most $4$ for computing each solution.




\section{Conclusions}
In this paper we present a new approach  for solving the portfolio selection
problem.
Instead of the standard Markowitz mean-variance model, we
have used an extension including buy-in threshold
and bounding constraints. These constraints make
the corresponding portfolio selection problem  nonconvex and so very difficult to solve
by existing algorithms. We have
transformed this problem into a mixed integer quadratic
program and  developed a deterministic approach based on DC programming and DCA.
Preliminary numerical simulations show the efficiency of DCA, its inexpensiveness and its superiority
with respect to standard branch-and-bound techniques.
They suggest to us extending the numerical experiments in higher dimension, and combining DCA and Branch and Bound algorithms for globally solving the problem of portfolio selection 
Work in these directions is currently in progress.


\end{document}